\documentclass[review]{elsarticle}
\usepackage[ruled,vlined]{algorithm2e}

\usepackage{booktabs}
\usepackage{multirow}

\usepackage{soul}
\usepackage{url}
\usepackage[utf8]{inputenc}
\usepackage[small]{caption}
\usepackage{graphicx}
\usepackage{amsmath}
\usepackage{booktabs}
\usepackage{multirow}
\usepackage{adjustbox}
\usepackage{enumitem}
\usepackage{amssymb}

\usepackage{xspace}
\newcommand{\mname}{MolTrans\xspace}

\usepackage{amsmath}
\usepackage{amsthm}

\theoremstyle{definition}

\newtheorem{problem}{Problem}

\journal{XXX}

\begin{document}

\begin{frontmatter}

\title{\mname: Molecular Interaction Transformer for Drug Target Interaction Prediction}

\author[first]{Kexin Huang}
\author[second]{Cao Xiao}
\author[second]{Lucas Glass}
\author[third]{Jimeng Sun}

\address[first]{Health Data Science, Harvard T.H. Chan School of Public Health, MA 02115}

\address[second]{Analytic Center of Excellence, IQVIA, MA 02139}

\address[third]{Department of Computer Science, University of Illinois at Urbana-Champaign, IL 61801}

\begin{abstract}
\textbf{Motivation:} Drug target interaction (DTI) prediction is a foundational task for in silico drug discovery, which is costly and time-consuming due to the need of experimental search over large drug compound space. Recent years have witnessed promising progress for deep learning in DTI predictions. However, the following challenges are still open: (1) the sole data-driven molecular representation learning approaches ignore the sub-structural nature of DTI, thus produce results that are less accurate and difficult to explain; (2) existing methods focus on limited labeled data while ignoring the value of massive unlabelled molecular data. \\
\textbf{Results:} We propose a \underline{Mol}ecular Interaction \underline{Trans}former (\mname) to address these limitations via: (1) knowledge inspired sub-structural pattern mining algorithm and interaction modeling module for more accurate and interpretable DTI prediction; (2) an augmented transformer encoder to better extract and capture the semantic relations among substructures extracted from massive unlabeled biomedical data. We evaluate \mname on real world data and show it improved DTI prediction performance compared to state-of-the-art baselines.\\
\textbf{Availability:} The model scripts and weights will be open sourced at Github.\\
\textbf{Contact:} jimeng@illinois.edu

\end{abstract}


\end{frontmatter}

\section{Introduction}

Drug discovery is notoriously costly and time-consuming due to the need of experimental search over large drug compound space. Take its key task, drug-target protein interaction (DTI) prediction as an example. DTI serves as the foundation for finding new drugs (i.e., virtual screening) and new indications of existing drugs (i.e., drug repositioning), since the therapeutic effects of drug compounds are detected by examining DTIs~\cite{hughes2011principles}. During the compound identification process, researchers often need to conduct assay experiments and search over 97M possible compounds in a candidate database~\cite{broach1996high}. 

Luckily, with massive biomedical data and knowledge being collected and available, along with the advances of deep learning technologies which demonstrated huge success in many application areas, the drug discovery process particularly DTI prediction has been significantly enhanced. Recently,  various deep models have shown encouraging performance in DTI predictions. They often take drug and protein data as inputs, cast DTI as a classification problem, and make prediction by feeding the inputs through deep learning models such as deep neural network (DNN)~\cite{unterthiner2014deep}, deep belief network (DBN)~\cite{wen2017deep}, and convolutional neural network (CNN)~\cite{mayr2018large,ozturk2019widedta,ozturk2018deepdta}. Despite these efforts, the following challenges are still open.

\begin{enumerate}[leftmargin=*]
    \item \textbf{Inadequate modeling of interaction mechanism}. 
     Existing works~\cite{ozturk2019widedta,ozturk2018deepdta,gao2018interpretable} learn molecular representation and make prediction based on whole molecular structure of drugs and proteins, ignoring that the interactions are sub-structural-- only involving relevant sub-structures of drugs and proteins~\cite{jia2009mechanisms,schenone2013target}. The full-structural molecular representations introduce noises and affect the prediction performance. Also, the learned representations are hard to interpret since they do not provide a tractable path to indicate which sub-structures of drugs and proteins contribute to the interactions. 
    \item \textbf{Restricted to limited labeled data}.
    Previous works~\cite{ozturk2019widedta,ozturk2018deepdta,gao2018interpretable,lee2019deepconv,wen2017deep} focus on data in hand and limit the scope within several thousands of drugs and proteins while ignoring the vast (e.g., order of millions) unlabelled biomedical data available. The model architectures in previous works also are not designed to enable the integration of massive dataset. 
\end{enumerate}
\noindent To solve these challenges, we propose  a transformer~\cite{vaswani2017attention}-based bio-inspired molecular data representation method (coined as \mname) to leverage vast unlabelled data for \textit{in silico} DTI prediction.

\mname is enabled by the following technical contributions:
\begin{enumerate}[leftmargin=*]
    \item \textbf{Knowledge inspired representation and interaction modeling for more accurate and explainable prediction}. Inspired by the knowledge that DTI is sub-structural, \mname~derives a data-driven method called Frequent Consecutive Sub-sequence (FCS) mining that is adaptable to extract high-quality fit-sized sub-structures for both protein and drug. In addition, \mname includes a bio-inspired interaction module imitating the real biological DTI process. The new sub-structure fingerprints enable a tractable path for understanding which sub-structure combination has more relevance to the outcome through an explicit map in the interaction module.
    
    \item \textbf{Leverage massive unlabelled biomedical data}. 
    \mname mines through millions of drugs and proteins sequences from multiple unlabelled data sources to extract high quality sub-structures of drugs and proteins. The vast data result in a much higher quality sub-structures than using small training dataset alone. We also augment the representation using transformers~\cite{vaswani2017attention}, which captures the complex signals among the large sequential sub-structures outputs generated from the unlabelled data.  

\end{enumerate}
We provide a comprehensive performance comparison of state-of-the-art methods on various realistic drug discovery settings include unseen drug/target problems and in scarce training dataset setup. We show empirically that \mname~has robust improved predictive performance over state-of-the-art baselines.

\section{Related Works}
Numerous computational methods have been developed for DTI prediction problem. Similarity-based methods such as kernel regression~\cite{pahikkala2014toward} and matrix factorization~\cite{zheng2013collaborative} methods exploit known DTI's drug-target similarity information and infer new ones. However, these methods are shown to be not generalizable to different protein classes~\cite{wen2017deep}. Feature-based methods feed numerical descriptors of drug and proteins into downstream prediction models. Popular numerical descriptors include ECFP~\cite{rogers2010extended} and PubChem~\cite{bolton2008pubchem} for drugs, CTD~\cite{dubchak1995prediction} and PSC~\cite{cao2013propy} for proteins. Classic machine learning methods such as gradient boosting~\cite{he2017simboost} have shown promises in predictive performance. Recently, deep learning based methods~\cite{unterthiner2014deep,wen2017deep,ozturk2019widedta,tsubaki2018compound} have shown further improvement of performance due to its capability to capture complex non-linear signals of DTI. \mname differs from existing works with (1) its knowledge-driven model architecture design rather than direct application of existing deep learning models; (2) emphasis on interpretability instead of predictive performance alone to potentially aid medical chemists for better decision making; (3) usage of external drug and target data to complement interaction dataset.   

\section{Method}

\subsection{Problem Definition}\label{section:problem} 
We formulate the DTI prediction as a classification task to determine whether a pair of drug and target protein will interact. In our setting, drug is represented by the Simplified Molecular Input Line Entry System (SMILES) $\mathbf{S}_i$, which consists of a sequence of chemical atoms and bonds tokens (e.g. C, O, S), generated by depth-first traversal over the molecule graph. We denote $\mathbf{S}$ for drug's SMILES representation. Target protein, denoted as $\mathbf{A}$, is represented by a sequence of protein tokens, where each token is one of the 23 amino acids. The DTI prediction task is defined as below.

\begin{problem}[\textbf{Drug Target Interaction (DTI) Prediction}]
Given compound sequence $\mathcal{S} = \{\mathbf{S}_1, \cdots, \mathbf{S}_n\}$ for $n$ drugs and protein sequence $\mathcal{A} = \{\mathbf{A}_1, \cdots, \mathbf{A}_m\}$ for $m$ proteins, the DTI prediction task can be casted as to learn a function mapping $\mathcal{F}: \mathcal{S} \times \mathcal{A} \rightarrow [0 ,1]$ from drug-target pairs to an interaction probability score.
\end{problem}

\begin{table}[t]
    \centering
    \caption{Main notations used in \mname.}
    \label{Notation Table}
    \begin{adjustbox}{max width=0.8\textwidth}

    \begin{tabular}{ll}\hline
    \bf Notations & \bf Description \\ \hline
    $\mathbf{S}, \mathbf{A}$ & drug SMILES, protein amino acids \\
    $\mathcal{I}, \mathcal{U}$ & the set of interacting/non-interacting DTI pairs \\ \hline
    $\mathcal{C}_p, \mathcal{C}_d$ & entire sub-structures set for protein and drug \\
    $\mathbf{C}_\mathrm{p}, \mathbf{C}_\mathrm{d}$ & sub-structures set in one pair of drug-target \\
    $\mathbf{M}^\mathrm{p} \in {\{0,1\}}^{k \times \Theta_p}$ & one-hot input representation for protein \\
    $\mathbf{M}^\mathrm{d} \in {\{0,1\}}^{j \times \Theta_d}$ & one-hot input representation for drug 
    \\ \hline
    $\mathbf{E}^\mathrm{p} \in \mathbb{R}^{\vartheta \times \Theta_p}$ & latent representation for protein \\
    $\mathbf{E}^\mathrm{d} \in \mathbb{R}^{\vartheta \times \Theta_d}$ & latent representation for drug  \\
    $\mathbf{I} \in \mathbb{R}^{k \times l \times \Psi} $ & the interaction tensor \\
    $\mathrm{F}$; $\mathbf{P} \in [0,1]$ & the interaction function; interaction probability \\
    $\mathbf{O} \in \mathbb{R}^{\varphi}$ & the output of interaction module \\
    $\mathbf{W}_\mathrm{cont}^\mathrm{p} \in \mathbb{R}^{\vartheta \times k}$ & the weight for content embedding protein \\
    $\mathbf{W}_\mathrm{cont}^\mathrm{d} \in \mathbb{R}^{\vartheta \times l}$ & the weight for content embedding drug  \\
    $\mathbf{W}_\mathrm{pos}^\mathrm{p} \in \mathbb{R}^{\vartheta \times \Theta_p}$ & the weight for position embedding protein \\
    $\mathbf{W}_\mathrm{pos}^\mathrm{d} \in \mathbb{R}^{\vartheta \times \Theta_d}$ & the weight for position embedding drug  \\
    $ \mathbf{W}_\mathrm{o} , \mathbf{b}_\mathrm{o} \in \mathbb{R}^{\varphi \times 1}$ & the weight, bias for decoder \\
    \hline 
    \end{tabular}
    \end{adjustbox}
\end{table}

\subsection{The \mname Method}
The \mname framework learns to predict DTI as follows.
Given the input drug and protein data, a \textbf{frequent consecutive sub-sequence mining module} first decomposes them into a set of explicit sequences of sub-structures using a specialized decomposition algorithm with inputs consisting of vast unlabelled data. The outputs are then fed into a \textbf{augmented transformer embedding module} to obtain an augmented contextual embedding for each sub-structure through transformer encoders~\cite{vaswani2017attention}. Next, in the \textbf{interaction prediction module},  drug sub-structures are paired with protein sub-structures with pairwise interaction scores. A CNN layer is later applied on the interaction map to capture higher-order interactions. Finally, a decoder module outputs a score indicating the probability of pairwise interactions. 

\begin{figure}[t]
    \centering
    \includegraphics[width =  \textwidth]{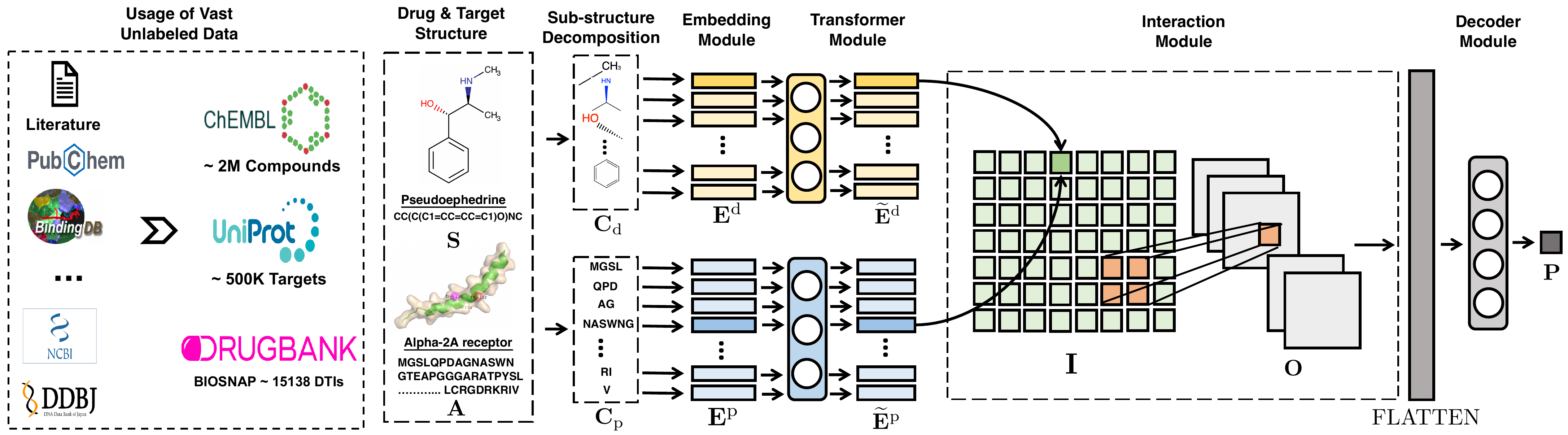}
    \caption{$\mname$ workflow: (a) $\mname$ utilizes vast unlabelled data. (b) Given the input pair of drug $\mathbf{S}$ and protein $\mathbf{A}$, $\mname$ extracts a sequence of sub-structures $\mathbf{C}_\mathrm{d}$ and $\mathbf{C}_\mathrm{p}$ via Algo. \ref{FCS}. (c) Each sub-structure is embedded into a latent feature vector $\mathbf{E}^\mathrm{d}$ and $\mathbf{E}^\mathrm{p}$ through a learnable embedding table via Eq.~\ref{eq:embed}. Then, drug/protein sequence of sub-structure embedding is fed into drug/target transformer encoders respectively to obtain an augmented contextual representation  $ \widetilde{\mathbf{E}}^\mathrm{d}$ and $\widetilde{\mathbf{E}}^\mathrm{p}$ via Eq.~\ref{eq:transformer}. (d) An interaction map $\mathbf{I}$ measuring interaction intensity among sub-structures is generated via Eq.~\ref{eq:interaction}. The interaction is further optimized by a CNN layer that models higher-order interaction, which results in a tensor $\mathbf{O}$ via Eq.~\ref{eq:densenet}. (e) A decoder module then feed the tensor for a classifier to output the DTI probability $\mathbf{P}$ via Eq.~\ref{eq:decoder}. All modules are trained end-to-end with the binary classification loss via Eq.~\ref{eq:loss}. 
    }
    \label{fig:mian}
\end{figure}

\subsubsection{Frequent Consecutive Sub-sequence Mining Module}
\begin{algorithm}[t] 
\textbf{Input:} $\mathbb{V}$ as the set of all initial amino acids/SMILES tokens; $\mathbb{W}$ as the set of tokenized proteins/drugs; $\theta$ as the specified frequency threshold; $\ell$ as the maximum size of $\mathbb{V}$. \\
\textbf{Output:} $\mathbb{W}$, the updated tokenized proteins/drugs; $\mathbb{V}$, the updated token vocabulary set. \\
\For{$t = 1 \ldots \ell$}{
     $\textsc{(A, B), freq} \leftarrow \mathrm{scan}\ \mathbb{W}$ \\
     //~ $\textsc{(A, B)}$ is the frequentest consecutive tokens. \\
    \If{$\textsc{freq} < \theta$}{$\mathrm{break}$ \hfill //~ $\textsc{(A, B)}$ \text{'s frequency lower than threshold}} 
    $\mathbb{W} \leftarrow \mathrm{find} \textsc{(A, B)} \in \mathbb{W},  \mathrm{replace}~\mathrm{with}\  \textsc{(AB)}$ \\
    //~\text{update $\mathbb{W}$ with the combined token \textsc{(AB)}} \\
    $\mathbb{V} \leftarrow \mathbb{V} \cup \textsc{(AB)}$ \hfill //~ \text{add \textsc{(AB)} to the token vocabulary set $\mathbb{V}$}
}

 \caption{Frequent Consecutive Sub-sequence Mining}
\label{FCS}
\end{algorithm}

Driven by the domain knowledge that DTI happens in a sub-structural level, \mname first decomposes molecular sequence for proteins and drugs into sub-structures. 
In particular, we propose a data-driven sequential pattern mining algorithm called Frequent Consecutive Sub-sequence Algorithm ($\mathrm{FCS}$) to find recurring sub-sequences across drug and protein databases. Inspired by the invention of sub-word units in the natural language processing field~\cite{sennrich2015neural,gage1994new}, $\mathrm{FCS}$ aims to generate a set of hierarchy of frequent sub-sequences for sequences. The algorithm is summarized in Algo. Using $\mathrm{FCS}$ algorithm, \mname converts input drug and target to a sequence of explicit sub-structures $\mathbf{C}_\mathrm{d}$ and $\mathbf{C}_\mathrm{p}$ respectively. The significance of $\mathrm{FCS}$ is two-folds:

\begin{enumerate}[leftmargin=*]
\item It distinguishes from previous sub-structure fingerprinting methods as it is more explainable. Commonly used fingerprint such as ECFP~\cite{rogers2010extended} is not explainable due to its hashing nature. Explicit sub-structure fingerprint such as PubChem encoding has on average 100 granular sub-structures for a small molecule where many sub-structures are a subset of other ones, making it intractable to know which sub-structure leads to the outcome. In contrast, FCS drug encoding is capable of giving explicit hints as it decomposes each drug molecule into discrete and moderate size \textit{partitions} of sub-structures as shown in Sec.~\ref{sec:explain}.
\item It allows for leveraging the massive unlabelled data for improved sub-structure mining. For example, we use the Uniprot dataset~\cite{boutet2007uniprotkb} consists of 560,823 unique protein sequences and the ChEMBL database~\cite{gaulton2011chembl} which includes 1,870,461 drug SMILES strings. We observe that the quality of the mined sub-structures originates from the massive unlabelled data we used. In small datasets, the occurrences of many useful sub-structures are below the reasonable minimum frequency whereas a large aggregation dataset can successfully identify them with a larger sequences pool. We also show that the encoding has better predictive power when using massive unlabelled data compared to using small datasets, in Sec.~\ref{sec:ablation}.
\end{enumerate}

\subsubsection{Augmented Transformer Embedding Module} To capture the chemical semantics of sub-structures, \mname includes an augmented embedding module where it first initializes a learn-able sub-structure look-up dictionary and then augment the embedding with the contextual sub-structural information via transformer encoders~\cite{vaswani2017attention}.

Concretely, for each input drug-target pair, we transform the corresponding  sequence of sub-structures $\mathbf{C}_\mathrm{p}$ and $\mathbf{C}_\mathrm{d}$ into two matrices $\mathbf{M}^\mathrm{p} \in \mathbb{R}^{k \times \Theta_p}$ and $\mathbf{M}^\mathrm{d} \in \mathbb{R}^{l \times \Theta_d}$ where $k/l$ is the total size of sub-structures for drug/protein or the cardinality of the vocabulary set $\mathbb{V}$ from FCS algorithm, $\Theta_p$ and $\Theta_d$ are the maximum lengths of sub-structure sequences for protein and drug, and each column $\mathbf{M}^\mathrm{p}_i$ and $\mathbf{M}^\mathrm{d}_j$ is a one-hot vector corresponding to the sub-structure index for the $i$-th sub-structure of protein sequence and $j$-th sub-structure of drug sequence. 
The content embedding ${\mathbf{E}_\mathrm{cont}^\mathrm{p}}_i, {\mathbf{E}_\mathrm{cont}^\mathrm{d}}_j$ for each protein and drug is generated via a learnable dictionary lookup matrix $\mathbf{W}_\mathrm{cont}^\mathrm{p} \in \mathbb{R}^{\vartheta \times k}$ and $\mathbf{W}_\mathrm{cont}^\mathrm{d} \in \mathbb{R}^{\vartheta \times l}$ such that
\begin{equation}
    {\mathbf{E}_\mathrm{cont}^\mathrm{p}}_i = \mathbf{W}_\mathrm{cont}^\mathrm{p} \mathbf{M}^\mathrm{p}_i, \hspace{2mm} {\mathbf{E}_\mathrm{cont}^\mathrm{d}}_j = \mathbf{W}_\mathrm{cont}^\mathrm{d} \mathbf{M}^\mathrm{d}_j,
\end{equation}
where $\vartheta$ is the size of latent embedding for each sub-structure.

Since \mname~uses sequential sub-structures, we also include a positional embedding ${\mathbf{E}^\mathrm{p}_\mathrm{pos}}_i, {\mathbf{E}^\mathrm{d}_\mathrm{pos}}_j$
via a lookup dictionary~\cite{vaswani2017attention} $\mathbf{W}^\mathrm{p}_\mathrm{pos} \in \mathbb{R}^{\vartheta \times \Theta_p}$ and $\mathbf{W}^\mathrm{d}_\mathrm{pos} \in \mathbb{R}^{\vartheta \times \Theta_d}$:
\begin{equation}
    {\mathbf{E}^\mathrm{p}_\mathrm{pos}}_i = \mathbf{W}^\mathrm{p}_\mathrm{pos} \mathbb{I}^\mathrm{p}_i, \hspace{2mm} {\mathbf{E}^\mathrm{d}_\mathrm{pos}}_j = \mathbf{W}^\mathrm{d}_\mathrm{pos} \mathbb{I}^\mathrm{d}_j,
\end{equation}
where $\mathbb{I}^\mathrm{p}_i \in \mathbb{R}^{\Theta_p}/\mathbb{I}^\mathrm{d}_j \in \mathbb{R}^{\Theta_d}$ is a single hot vector where $i/j$th position is 1. 

The final embedding $\mathbf{E}^\mathrm{p}_i, \mathbf{E}^\mathrm{d}_j$ are generated via the sum of content and positional embedding:
\begin{equation}\label{eq:embed}
    \mathbf{E}^\mathrm{p}_i = {\mathbf{E}_\mathrm{cont}^\mathrm{p}}_i + {\mathbf{E}^\mathrm{p}_\mathrm{pos}}_i, \hspace{2mm} \mathbf{E}^\mathrm{d}_j = {\mathbf{E}_\mathrm{cont}^\mathrm{d}}_j + {\mathbf{E}^\mathrm{d}_\mathrm{pos}}_j.
\end{equation}

The models above outputs a set of independent sub-structure embedding. However, these sub-structures have chemical relationships (e.g. Octet rules) among themselves to capture these contextual information, we further augment the embedding using a transformer encoder layers~\cite{vaswani2017attention}:
\begin{equation}\label{eq:transformer}
    \widetilde{\mathbf{E}}^\mathrm{p} = \mathrm{Transformer}_\mathrm{Protein}(\mathbf{E}^\mathrm{p}), \hspace{2mm}
    \widetilde{\mathbf{E}}^\mathrm{d} = \mathrm{Transformer}_\mathrm{Drug}(\mathbf{E}^\mathrm{d})
\end{equation}
Transformer encoder is ideal in our case because in the core of a transformer encoder, it is a series of self-attention head that modify each individual sub-structure embedding by learning the distribution of the effects from all the other sub-structures. 

\subsubsection{Interaction Prediction Module}\label{section:interaction}
\mname~includes an interaction module that consists of two layers: (a) an interaction tensor to model pair-wise sub-structural interaction; (b) a CNN layer over interaction map to extract neighborhood interaction.\\

\noindent\textbf{Pairwise interaction}. To model the pairwise interaction, for each sub-sequence $i$ in protein and sub-sequence $j$ in drug, we have
\begin{equation}\label{eq:interaction}
    \mathbf{I}_{\mathrm{i,j}} = \mathrm{F}(\widetilde{\mathbf{E}}^\mathrm{p}_\mathrm{i}, \widetilde{\mathbf{E}}^\mathrm{d}_\mathrm{j}), 
\end{equation}
where $\mathrm{F}$ is a function that measures the interaction between the pairs. It can be any function such as sum, average, and dot product. Therefore, after this layer, we have a tensor $\mathbf{I} \in \mathbb{R}^{\Theta_d \times \Theta_p \times \Phi}$ where $\Theta_d/\Theta_p$ is the length of sub-sequences for drug/protein respectively, and $\Phi$ is the size of the output of function $\mathrm{F}$, where  Each column in this tensor takes account into the interaction of individual sub-structure of proteins and drugs. To provide explainability, we favor dot product as the aggregation function because it generates a single scalar that explicitly measures the intensity of interaction between individual target-drug sub-structural pair. As dot product output is one dimensional for every pair, $\mathbf{I}$ becomes a two-dimensional interaction map. By examining this map, we directly see which sub-structure pairs contribute to the final outcome. \\

\noindent\textbf{Neighborhood interaction}. Nearby sub-structure of proteins and drugs influence each other in triggering the interactions. Hence, besides modeling the individual pair-wise interaction, it is also necessary to model the interaction to the nearby regions. We achieve this through a Convolutional Neural Network (CNN)~\cite{krizhevsky2012imagenet} layer on top of the interaction map $\mathbf{I}$. The intuition is that by applying several order-invariant local convolution filters, interaction to nearby regions can be captured and aggregated. We obtain the output representation $\mathbf{O}$ of the input drug-target pair: 
\begin{equation}\label{eq:densenet}
    \mathbf{O} = \mathrm{CNN}(\mathbf{I})
\end{equation}

This interaction module is inspired from the Deep Interactive Inference Network~\cite{gong2017natural}. Thanks to this explicit interaction modeling, we can later visualize the strength of individual sub-structural interaction pair from the interaction map. To output a probability indicating the likelihood of interaction, we first flatten the $\mathbf{O}$ into a vector and use a linear layer parametrized by weight matrix $\mathbf{W}_\mathrm{o}$ and bias vector $\mathbf{b}_\mathrm{o}$:
\begin{equation}\label{eq:decoder}
    \mathbf{P} = \sigma ( \mathbf{W}_\mathrm{o} \mathrm{FLATTEN}(\mathbf{O}) + \mathbf{b}_\mathrm{o} ),
\end{equation}
where $\sigma(a) = \frac{1}{1 + \exp{(-a)}}$.

The entire network with parameters $\mathbf{W}_\mathrm{cont}^\mathrm{p}$, $\mathbf{W}_\mathrm{cont}^\mathrm{d}$, $\mathbf{W}_\mathrm{pos}^\mathrm{p}$, $\mathbf{W}_\mathrm{pos}^\mathrm{d}$ $\mathbf{W}_\mathrm{o}$, $\mathbf{b}_\mathrm{o}$, the Transformer Encoders weights and CNN weights can be jointly optimized through the binary classification loss:
\begin{equation}\label{eq:loss}
    \mathbf{L} = \mathbf{Y}~\mathrm{log}~(\mathbf{P}) + (1 - \mathbf{Y})~\mathrm{log}~(1 - \mathbf{P}),
\end{equation}
where $\mathbf{Y}$ is the ground truth label.

\section{Experiment}
We design experiments to answer the following questions.

\begin{enumerate}[leftmargin=*]
    \item[\textbf{Q1}]: Does \mname~improve DTI predictive performance?
    \item[\textbf{Q2}]: How well does \mname~tackle the unseen-drug/target cases? 
    \item[\textbf{Q3}]: How does \mname respond to large number of missing data?
    \item[\textbf{Q4}]: Does \mname~provide useful knowledge about DTI?
    \item[\textbf{Q5}]: How does each component of \mname contribute to the predictive performance gain?
\end{enumerate}

\subsection{Experimental Setup}
\noindent \textbf{Dataset.} We use the MINER DTI dataset from BIOSNAP collection~\cite{biosnapnets}. It consists of 4,503 drug nodes and 2,182 protein targets, and 15,138 drug-target interaction pairs from DrugBank~\cite{drugbank}. BIOSNAP dataset only contains positive drug target interaction pairs. For negative pairs, we sample from the unseen pairs. We obtain a balanced dataset with equal positive and negative samples. We use Uniprot database to obtain target sequence and DrugBank database to obtain SMILES strings, and hence, we filter out target/drug that do not have target sequences/SMILES from Uniprot/DrugBank. \\

\noindent{\bf Metrics.} We use ROC-AUC (Area under the receiver operating characteristic curve) and PR-AUC (Area under the precision-recall curve) as metrics to measure the binary classification performance.

\noindent \textbf{Evaluation Strategies.} We divided the dataset into training, validation and testing sets in a 7:1:2 ratio. For every experiment, we conduct five independent runs with different random splits of dataset. We then select the best performing model based on ROC-AUC performance from the validation set. The selected model via validation is then evaluated on the test set with the result reported below.

\noindent \textbf{Implementation Details.} \mname is implemented in PyTorch~\cite{paszke2017automatic}. For experiments, we use a server with 2 Intel Xeon E5-2670v2 2.5GHZ CPUs, 128GB RAM and 2 NVIDIA Tesla P40 GPUs. For \mname, we use Adam optimizer with learning rate of 1e-5. We set the batch size to be 64 and we allow it to run for 30 epochs. It converges between 8-15 epochs. The input embedding is of size 384 and we set 12 attention heads for each transformer encoder with intermediate dimension 1,536. The dropout rate is 0.1. We set the maximum length of sequence for both drug and protein to cover 95$\%$ of drugs and proteins in the dataset. For the CNN, we use three filters with kernel size three. 

\begin{table}[t]
    \centering
    \caption{Performance comparison (5 random runs).}
    \begin{tabular}{lcc}
    \toprule
    Method & ROC-AUC & PR-AUC \\ \hline
    KronRLS & $0.841 \pm 0.005$ & $0.810 \pm 0.009$ \\
    LR & $0.846 \pm 0.004$ & $0.850 \pm 0.011$ \\
    DNN & $0.849 \pm 0.003$ & $0.855 \pm 0.010$ \\
    DeepDTI & $0.876 \pm 0.005$ &$0.876 \pm 0.006$\\
    DeepDTA & $0.876 \pm 0.005$ & $0.883 \pm 0.006$\\
    GNN-CPI & $0.879 \pm 0.007$ & $0.890 \pm 0.004$\\ 
    DeepConv-DTI & $0.883 \pm 0.002$ & $0.889 \pm 0.005$\\
    \mname & $\boldsymbol{0.895 \pm 0.002}$ & $ \boldsymbol{0.901 \pm 0.004}$  \\
    \bottomrule
    \end{tabular}
    \label{tab:q2}
\end{table}

\subsection{Baselines}

We compared \mname with the following baselines. We focus on state-of-the-art deep learning models as they have demonstrated superior performance over shallow models. 
\begin{enumerate}[leftmargin=*]
    \item \textbf{KronRLS}~\cite{pahikkala2014toward} applies a kernel least square method on the similarity matrix of drug and protein features. 
    \item \textbf{LR}~\cite{cao2013propy,rogers2010extended} applies a logistic regression model on the concatenated drug and protein feature vectors. We experiment on all the combinations for ECFP4~\cite{rogers2010extended} \& PubChem~\cite{wang2009pubchem} for drugs and protein sequence composition descriptor (PSC)~\cite{cao2013propy} \& Composition-Transition-Distribution (CTD)~\cite{dubchak1995prediction} for proteins. We find ECFP4 for drugs and PSC for protein has the highest performance. 
    \item \textbf{DNN} uses a three layer deep neural network with hidden size 1,024 on top of the ECFP4 and PSC concatenated vector. 
    \item \textbf{DeepDTI}~\cite{wen2017deep} models DTI using Deep Belief Network~\cite{hinton2009deep}, which is a stack of Restricted Boltzmann Machines~\cite{hinton2012practical}. It uses the concatenation of ECFP2, ECFP4, ECFP6 as the drug feature and uses PSC for protein features. We optimize the hyper-parameters described from the paper based on validation set performance.  
    \item \textbf{DeepDTA}~\cite{ozturk2018deepdta} applies CNN on both raw SMILES string and protein sequence to extract local residue patterns. The task is to predict binding affinity values. We add a Sigmoid activation function in the end to change it to a binary classification problem and we conduct hyper-parameter search to ensure fairness. 
    \item \textbf{GNN-CPI}~\cite{tsubaki2018compound} uses graph neural network to encode drugs and use CNN to encode proteins. The latent vectors are then concatenated into a neural network for compound-protein interaction prediction. We follow the same hyper-parameter setting described in the paper.
    \item \textbf{DeepConv-DTI}~\cite{lee2019deepconv} uses CNN and global max pooling layer to extract various length local pattern in protein sequence and applies fully connected layer on drug fingerprint ECFP4. It conducts extensive experiment on different datasets and is the state-of-the-art model in DTI binary prediction task. We follow the same hyper-parameter setting described in the paper.
\end{enumerate}

\subsection{Q1: \mname achieves superior predictive performance}\label{sec:result}

To answer \textbf{Q1}, we randomly select 20$\%$ drug protein pairs as test set.~Table.~\ref{tab:q2} shows \mname has better predictive baselines in the DTI prediction setting in both metrics.

\subsection{Q2: \mname has competitive performance in unseen drug and target setting}

\begin{figure}[t]
    \centering
    \includegraphics[width = 0.6\textwidth]{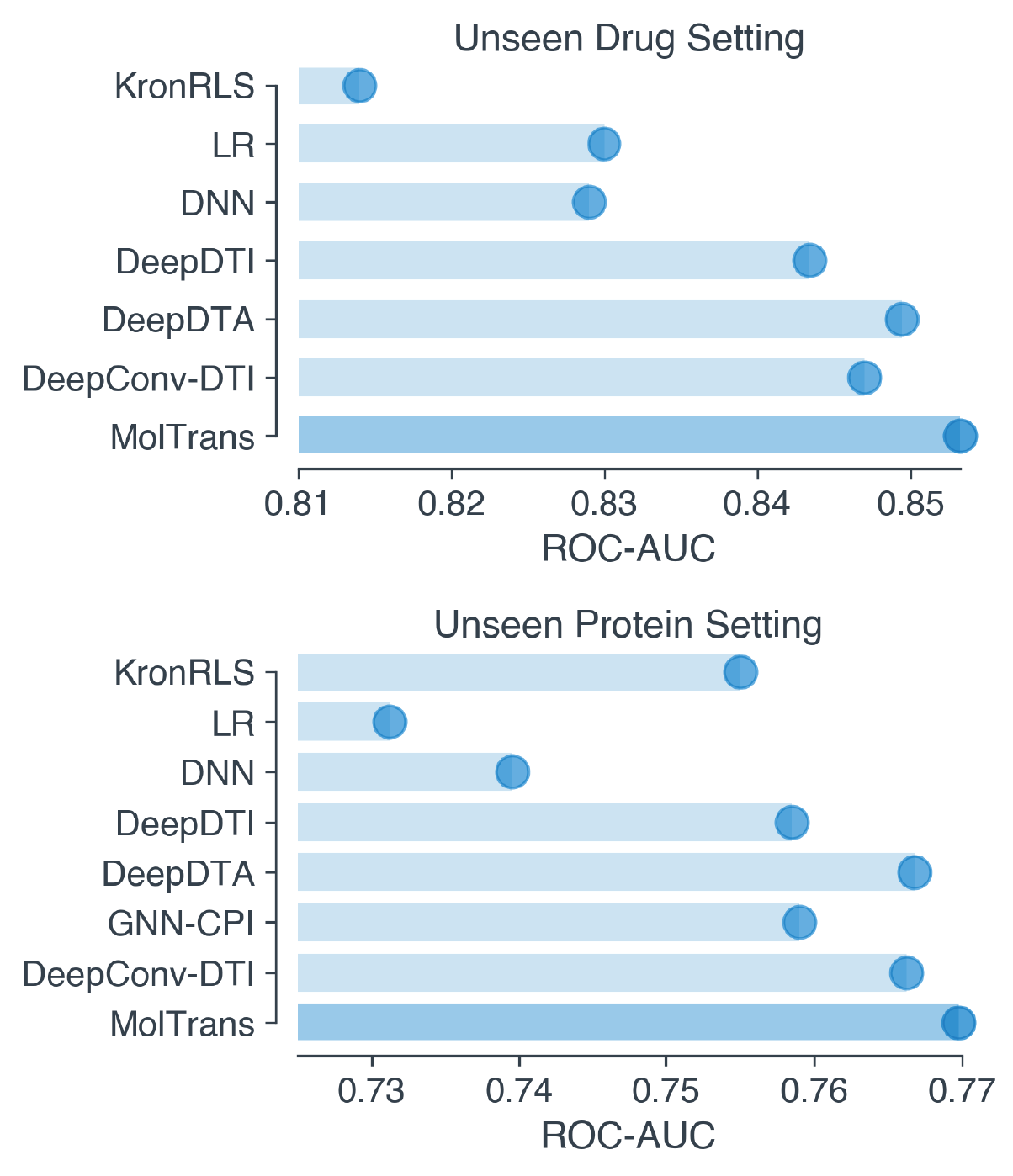}
    \caption{\mname has competitive result in both unseen drug and protein settings (shown avg. of 5 random runs). We omit GNN-CPI in unseen drug setting due to its bad performance (0.749). 
    }
    \label{fig:unseen}
\end{figure}

To imitate the unseen drug/target task, we randomly select 20\% drug/target proteins and all DTI pairs associated with these drugs and targets as the test set. 

The results are in Fig.~\ref{fig:unseen}. We observe that KronRLS's performance vary across settings. This is because KronRLS is a similarity-based method, hence it is susceptible to the data properties in hand. In the unseen drug setting, we find the one-layer LR is better than multi-layers DNN, and is worse than the SOTA methods with more complicated deep model design. This shows the necessity for carefully designed model architecture. We also see that \mname has competitive performance against the SOTA deep learning baselines in both settings. 

\subsection{Q3: \mname performs best with scarce data}
\begin{figure}[h]
    \centering
    \includegraphics[width=0.7\textwidth]{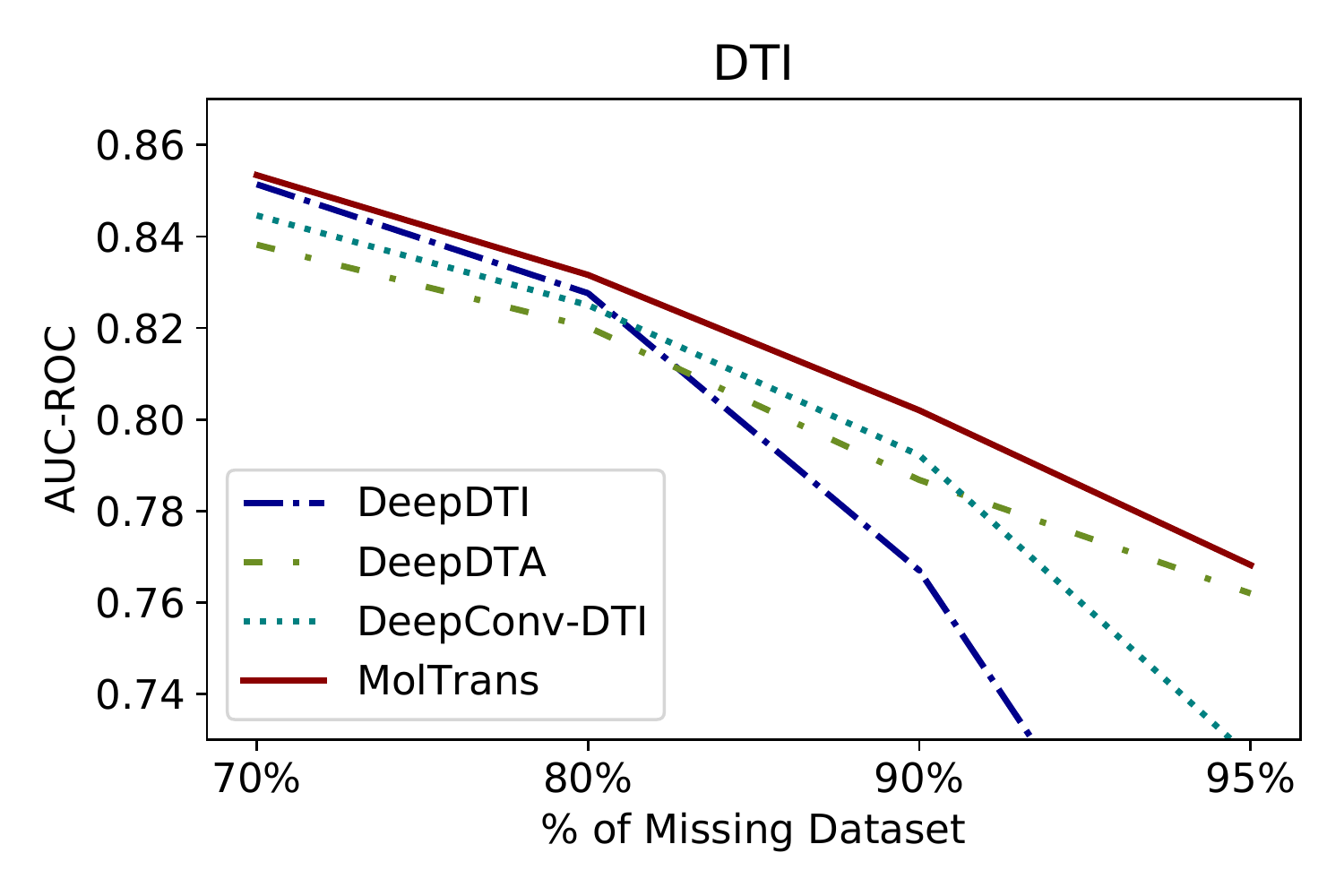}
    \caption{\mname provides best result in high fraction of missing data (shown avg. of 5 random runs). GNN-CPI is worse than all the other methods and is thus omitted.}
    \label{fig:missing}
\end{figure}

Robust performance under scarce training dataset is ideal in DTI setting. We trained each method on 5\%, 10\%, 20\%, and 30\% of dataset and predict on the rest of them (we use 10\% of the test edges as validation set for early stopping). The result is reported in Fig.~\ref{fig:missing}. We see that \mname is the most robust method. In the contrast, SOTA baselines such as DeepDTI and DeepConv-DTI drop as missing fractions increase. One reason why \mname is good on scarce setting is that \mname leverages on embeddings from sub-structures which are relatively abundant hence transferable compared to other methods which utilize the entire drugs and proteins.

\subsection{Q4: \mname~allows model understanding}\label{sec:explain}
\begin{figure}[t]
    \centering
    \includegraphics[width = 0.8\textwidth]{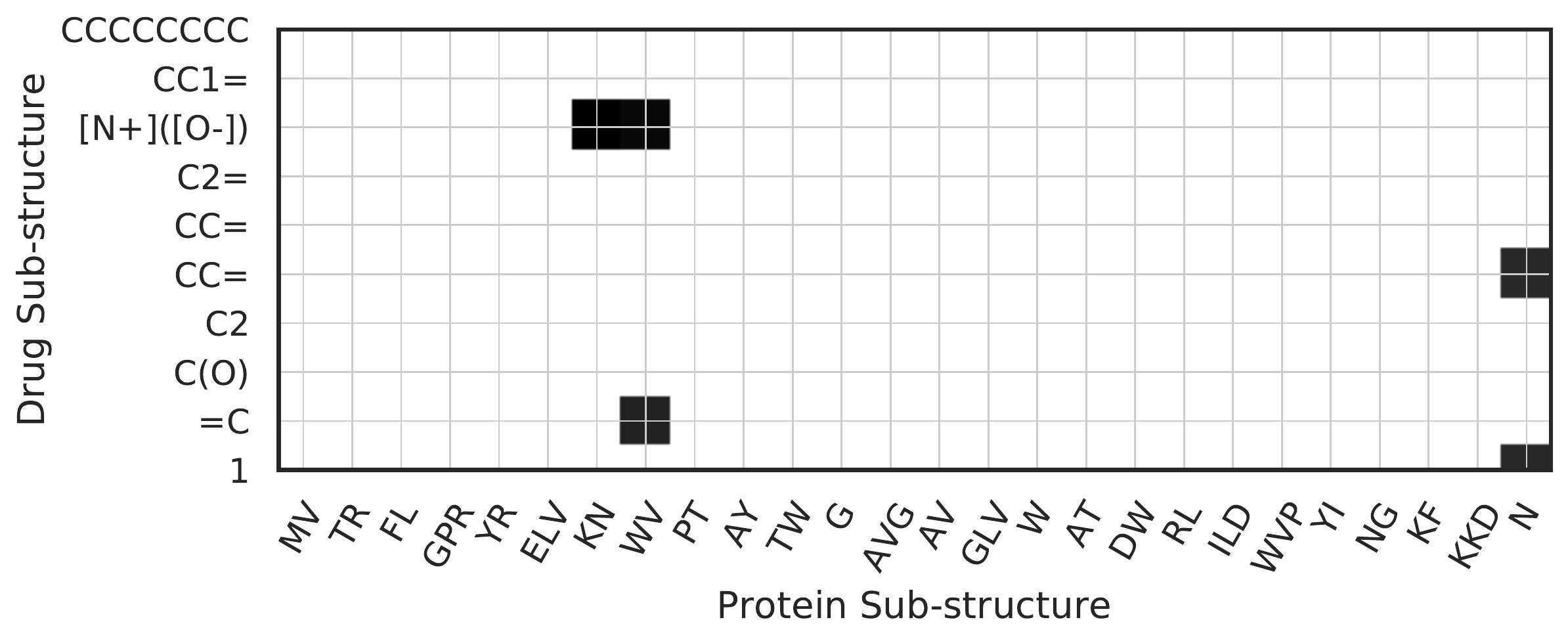}
    \caption{The interaction map on the contributions of sub-structures in DTI, shown as drug 2-nonyl n-oxide interacts with protein Cytochrome b-c1 complex unit 10. 
    }
    \label{fig:2}
\end{figure}

To answer \textbf{Q4}, we show through examples how the interaction map $\mathbf{I}$ can provide hints on which sub-structure leads to the interaction. We first feed drug 2-nonyl n-oxide, and the protein Cytochrome b-c1 complex unit 1 into \mname, and we visualize the interaction map by filtering scalars that are larger than a threshold in Fig.~\ref{fig:2}. We saw the nitrogen oxide group [N+]([O-]) and KNWV has the highest interaction coefficient, matching with the previous study~\cite{lightbown1956inhibition} who showed that nitrogen oxide group is essential for cytochrome inhibition activity. This example supports that \mname~is capable of providing reasonable cues for understanding the model prediction and possibly shed light on the inner workings of DTI. To add more credibility, we feed Ephrin type-A receptor 4 (Epha4) target and Dasatinib drug into \mname, the map shows amino-thiazole group (S(=O)(=O) and N sub-structures) is highlighted with protein motif KF and DVG, which has an overlap with the Epha4-Dasatinib complex described in previous study~\cite{farenc2011crystal}. We also feed the input target protein HDAC2 and the input drug hydroxamic acid. The interaction map assigns the NC(=O) group and the carbon chain with protein sub-structure KK, YG, DIG, DD with high intensity. The suggested ligand sub-structure matches with the observed interaction in HDAC2-SAHA co-complex~\cite{lauffer2013histone}. 

\subsection{Q5: Ablation Study}\label{sec:ablation}
We conduct an ablation study on the full data setting with the following setup:

\begin{table}[t]
    \centering    
    \caption{Ablation study (5 random runs)}
    \begin{tabular}{lcc}
    \toprule
    Setup & ROC-AUC & PR-AUC \\ \hline
    \mname & $\boldsymbol{0.895 \pm 0.002}$ & $ \boldsymbol{0.901 \pm 0.004}$ \\ \hline
    -CNN & $0.876 \pm 0.003$ & $0.883 \pm 0.006$  \\
    -AugEmbed & $0.876 \pm 0.004$ & $0.870 \pm 0.004$ \\ 
    -Interaction & $0.847 \pm 0.003$ & $0.859 \pm 0.005 $ \\ \hline
    Small & $0.888 \pm 0.001$& $0.888 \pm 0.007$ \\
    -FCS & $0.887 \pm 0.004 $ & $0.887 \pm 0.004$ \\
    \bottomrule
    \end{tabular}
    \label{tab:ablation}
\end{table}

\begin{enumerate}[leftmargin=*]
    \item -CNN: we remove the CNN from interaction module, and flatten the interaction map $\mathbf{I}$ output and feed into the decoder.
    \item -AugEmbed: we remove the transformer in the augmented embedding module and feed the interaction module with the positional and content embedding,
    \item -Interaction: we further remove the interaction module from -AugEmbed. It degenerates to a decoder on top of the FCS fingerprint. Note that removing the interaction module alone is not a valid model design. 
    \item Small: we use smaller dataset to train FCS: DrugBank for drug and BindingDB for protein. We adjust the minimal frequency to output a similar number of sub-structured as FCS-large.
    \item -FCS: we replace FCS with other popular drug/protein encoding. Here, we use ECFP4 for drug and PSC for protein. 
\end{enumerate}

From Table.~\ref{tab:ablation}, we see CNN, transformers and interaction module contribute to the model final performance. The FCS fingerprint alone has strong predictive performance already from -Interaction. In addition, from Small, we see the massive unlabelled data is useful as it enriches the input and boosts the performance. From -FCS, we see our model is adaptable to other popular fingerprints with similar strong performance (better than all SOTA baselines).

\section{Conclusion}
In this work, we introduce \mname, an end-to-end biological inspired deep learning based framework that models DTI process. We test under realistic drug discovery setting and evaluate with state-of-the-art baselines. We demonstrate empirically that \mname has competitive performance in accurately predicting DTI under all settings with an improved explainability. 

\bibliography{mybibfile}
\end{document}